\documentclass[prd,aps,twocolumn]{revtex4}
\usepackage{amsmath}
\usepackage{graphicx}
\usepackage{amssymb}
\usepackage{hyperref}
\usepackage{color}

\def\P{{\cal P}}

\definecolor{verde}{rgb}{0,0.5,0}

\def\be{\begin{equation}}
\def\ee{\end{equation}}
\def\bea{\begin{eqnarray}}
\def\eea{\end{eqnarray}}
\def\be{\begin{equation}}
\def\ee{\end{equation}}
\def\ba{\begin{align}}
\def\ea{\end{align}}

\newcommand\lsim{\mathrel{\rlap{\lower4pt\hbox{\hskip0.5pt$\sim$}}
    \raise1pt\hbox{$<$}}}
\newcommand\gsim{\mathrel{\rlap{\lower4pt\hbox{\hskip0.5pt$\sim$}}
    \raise1pt\hbox{$>$}}}

\begin{document}

\renewcommand{\topfraction}{0.99}
\renewcommand{\bottomfraction}{0.99}

\title{A novel PBH production mechanism from non-Abelian gauge fields during inflation}

\author{Ema Dimastrogiovanni$^a$, Matteo Fasiello$^{b}$ and Alexandros Papageorgiou$^{b\,*}$}

\affiliation{$^a$Van Swinderen Institute for Particle Physics and Gravity, University of Groningen, Nijenborgh 4, 9747 AG Groningen, The Netherlands}
\affiliation{$^b$Instituto de Física Téorica UAM/CSIC, Calle Nicolás Cabrera 13-15, Cantoblanco, 28049, Madrid, Spain}
\email{alexandros.papageorgiou@ift.csic.es}

\begin{abstract}
We consider the case of axion-like particles (ALPs) during inflation. When coupled to a non-Abelian gauge sector via a Chern-Simons term, ALPs  support an intriguing, testable, phenomenology with very distinctive features including chiral primordial gravitational waves. For sufficiently small values of the gauge vev and coupling, scalar perturbations in the gauge sector exhibit a known instability. We harness the power of such instability for primordial black hole (PBH) generation.
In the case of an axion-inflaton, one is dynamically driven into a strong-backreaction regime that crosses the instability band thereby sourcing a peaked scalar spectrum leading to PBH production and the related scalar-induced gravitational waves.
Remarkably, this dynamics is largely insensitive to the initial conditions and the shape of the potential, highlighting the universal nature of the sourcing mechanism. In the case of spectator ALPs one can identify the parameter space that sets off the strong backreaction regime and the ensuing features. We show that spectator ALP models may also access the scalar instability region without triggering strong backreaction.
 \end{abstract}

\maketitle

\section{Introduction}
\label{intro}
The (first direct) detection of gravitational waves from the merger of two black holes in 2015 \cite{LIGOScientific:2016aoc} has brought about a strong, renewed, interest in PBH physics as these might be a compelling dark matter candidate \cite{Bird:2016dcv,Clesse:2016vqa,Sasaki:2016jop} (see \cite{Sasaki:2018dmp,LISACosmologyWorkingGroup:2023njw,Ozsoy:2023ryl} for recent reviews). One should stress the tantalizing possibility that PBH constitute all of the dark matter, an hypothesis that will be put to the test by interferometers such as LISA \cite{Bartolo:2018evs,Bartolo:2018rku}. It is in this context that we put forward a new  mechanism for PBH production in axion inflation. 

The use of axions and axion like particles (ALPs) has been ubiquitous in particle physics and cosmology since the idea was first put forward as a solution to the strong CP problem \cite{Peccei:1977hh,Peccei:1977ur,Weinberg:1977ma,Wilczek:1977pj}. ALPs make for a compelling presence also in the Lagrangian describing cosmic inflation \cite{Freese:1990rb,Adams:1992bn,Pajer:2013fsa}.~The (approximate) shift-symmetry of, for example, an axion-inflaton potential protects the inflaton mass from large quantum corrections thus addressing the so-called $\eta$-problem.~Phenomenological approaches to axion-inflation are further supported by the fact that axions are a generic prediction of string theory \cite{Witten:1984dg,Svrcek:2006yi,Douglas:2006es,Grimm:2012yq,Hebecker:2015tzo,Hebecker:2018yxs,Carta:2021uwv,Cicoli:2023opf}. Indeed, the intriguing notion of a string axiverse has been emerging in this context \cite{Arvanitaki:2009fg,Acharya:2010zx,Cicoli:2012sz,Demirtas:2018akl,Demirtas:2021gsq,Dimastrogiovanni:2023juq}.\\
The simplest model of axion inflation is a single-field setup known as natural inflation \cite{Freese:1990rb,Adams:1992bn}. A trans-Planckian axion decay constant $f$ has been necessary \cite{Freese:2014nla} to flatten its inflationary potential and ease the tension between the theory predictions and data from CMB observations \footnote{On account of the most recent data from Planck and BICAP/Keck \cite{Planck:2018jri,BICEP:2021xfz}, the model, at least in its simplest realization, has been ruled out.}. Such a large value is called into question by the expectation that at Planck scales the shift symmetry, as well as all global symmetries, is broken by quantum gravity effects \cite{Banks:2003sx}. This is not true if the  symmetry arises  from a gauge symmetry, such as is the case in string theory \cite{Baumann:2014nda}. However, string theory constructions typically deliver a sub-Planckian $f$. From these considerations stem the effort to deliver a viable cosmology for small $f$ values.\\ Several mechanisms have been explored, from multiple interacting axions \cite{Kim:2004rp} to dissipating the rolling axion-inflaton kinetic energy into a gauge sector via Chern-Simons (CS) coupling \cite{Anber:2009ua,Adshead:2012kp}. The latter scenario supports rather interesting and very distinctive features such as a chiral gravitational wave spectrum.~The rich phenomenology of this type of models and the exciting opportunity of testing them across the scales in the near future  \cite{Thorne:2017jft,Campeti:2020xwn,Smith:2016jqs,Domcke:2019zls} had led to a flurry of research activity on both Abelian and non-Abelian configurations. 
We shall focus on the latter class as it is the one supporting the key PBH production mechanism we want to highlight.\\
Some of the main features of the non-Abelian  case vis-\`{a}-vis its Abelian counterpart are (i) the linear sourcing of gravitational waves, (ii) the existence of an isotropic attractor \cite{Maleknejad:2013npa,Wolfson:2020fqz} solution for the background, and (iii) the possibility of slowing down the inflaton within the weak backreaction \footnote{The backreaction of gauge field fluctuations on the background.} regime.\\
For sufficiently small values of the gauge field coupling and background, scalar fluctuations in this sector display an instability \cite{Dimastrogiovanni:2012ew,Adshead:2013nka}. The parameter space corresponding to the instability and, until very recently \cite{Iarygina:2023mtj}, the strong-backreaction regime have not been investigated in the literature on non-Abelian theories, with the exception of \cite{Ishiwata:2021yne} which treated the backreaction terms analytically.~Spurred by the study in \cite{Iarygina:2023mtj}, in this work we explore both and uncover rather interesting findings.~The analysis of \cite{Iarygina:2023mtj} revealed that, if the strong backreaction regime is accessed, the background evolution crosses and eventually exits the scalar instability band towards an attractor solution.~The scalar instability may therefore be a finite one.~It behooves us then to address the following questions. How does the instability affect scalar fluctuations ?~Can it lead to  PBH production? If so, can one evade the latest PBH bounds?\\
We first (positively!) answer these questions in the case of an axion-inflaton. There the rolling down the potential will generically trigger the onset of strong backreaction and  evolve the system towards the attractor solution.~Such dynamics is remarkably insensitive to the initial conditions and the details of the inflationary potential.\\
The case of a spectator axion is more subtle. The potential for a spectator field is not bound to support (and end) the inflationary expansion so that one may now  evolve the background without ever accessing strong-backreaction. Crucially, this does not prevent the background from entering the scalar instability band, but rather it makes triggering the instability more dependent on the shape of the axion potential and the parameter space region being probed.\\ This paper is organized as follows. In Sections~\ref{section2} and \ref{crossing} we focus on the case of an axion-inflaton. ~We show how a peak in the curvature spectrum emerges rather generically during the backreaction regime.~We explore how this may result in primordial black hole production as well as in scalar-induced gravitational waves (SIGW). In Section~\ref{spectator} we turn to the spectator case and, in order to highlight a qualitatively different behaviour with respect to Sections~\ref{section2}-\ref{crossing}, we focus on the regime where strong backreaction is not encountered throughout the evolution. We show how the scalar instability may nevertheless still be accessed and used towards PBH production and the corresponding SIGW background. In Section~\ref{conclusions} we summarize and discuss our findings as well as point to natural follow-up work. Additional details on the analysis of the perturbations can be found in the supplemental material provided in the Appendices.

\section{Non-Abelian Models and the Scalar Instability Band}
\label{section2}
\begin{figure}
    \centering
    \includegraphics[width=0.99\linewidth,angle=0]{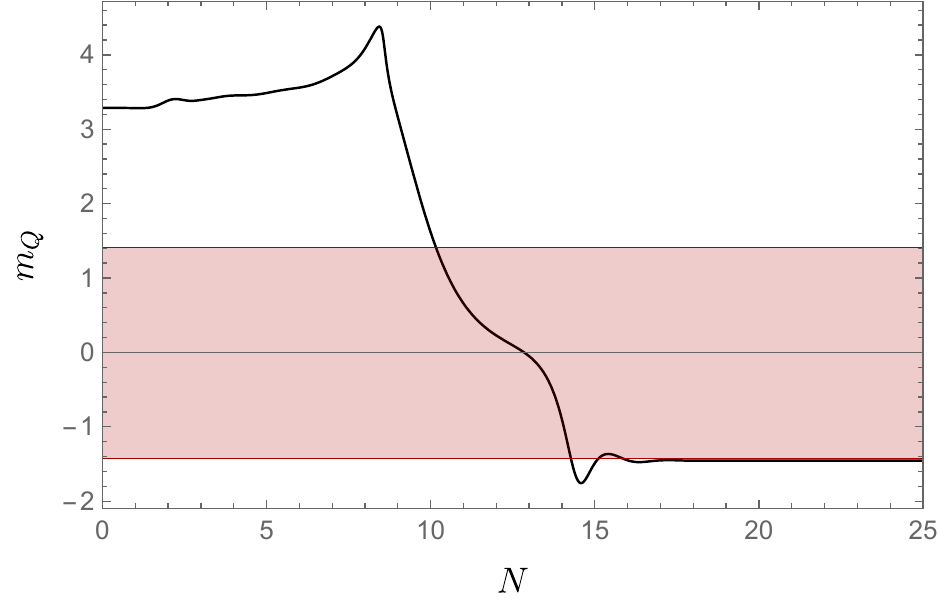}
    \caption{An example of the evolution of the particle production parameter with the full tensor backreaction terms included. The red shaded region corresponds to the well known $|m_Q|<\sqrt{2}$ instability band. We have verified the results of \cite{Iarygina:2023mtj} for the run $\mu3$ of the same paper using our own independently written code in \textit{Julia}.}
    \label{mQold}
\end{figure}

Let us consider the simplest inflationary field content that exhibits the mechanism we want to study. It consists of  an axion-inflaton coupled via a Chern-Simons term to a non-Abelian gauge sector (specifically an SU(2), hence the ``chromo-natural'' inflation nomenclature)  \cite{Adshead:2012kp}:
\bea
\begin{aligned}
&S=\int d^4 x \sqrt{-g}\Bigg[\frac{M_P^2}{2} R -\frac{1}{2}(\partial \chi)^2 - V(\chi) \\&\qquad\qquad\qquad\quad - \frac{1}{4}F_{\mu\nu}^a F^{a\,\mu\nu}+\frac{\lambda\,\chi}{4 f}F^{a}_{\mu\nu}   \tilde{F}^{a\,\mu\nu}\Bigg]\;,
\end{aligned}
\label{Lag}
\eea
where $F_{\mu\nu}^{a}\equiv\partial_{\mu}A^{a}_{\nu}-\partial_{\nu}A^{a}_{\mu}-g \epsilon^{abc} A^{b}_{\mu}A_{\nu}^{c}$ and $\tilde{F}$ is contracted with the fully antisymmetric tensor $\epsilon^{\mu\nu\rho\sigma}/\sqrt{-g}$. 
A simple counting argument \footnote{One can use the gauge freedom to fix 2 of the 4 entries in $A_{\mu}$ and, upon remembering there are 3 independent generators of SU(2), one obtains for $A^a_{\mu}$ a total of 6 propagating degrees of freedom.} suggests $2 \times 3$ degrees of freedom in the gauge sector: 2 traceless transverse tensors, two transverse vectors and 2 scalars. The non-Abelian nature of the gauge sector allows for an isotropic background solution:
\bea
A_0^a=0\quad, \quad A_{i}^a=\delta_i^a\,a(t)\,Q(t) \; ,
\eea
and we define $m_Q\equiv gQ/H$ and $\Lambda\equiv\lambda Q/f$ for later convenience.\\ Gauge field fluctuations are written explicitly in Appendix \ref{app.A}, covering both the axion-inflaton and axion-spectator case. The theory comprises three scalar degrees of freedom including the axion-inflaton. These modes are coupled to each other so that their dynamics is described by a non-diagonal $3\times3$ matrix. Crucially, the scalars in the gauge sector undergo an exponential growth whenever the condition $|m_Q|<\sqrt{2}$ is satisfied \cite{Dimastrogiovanni:2012ew}, an enhancement which is transmitted to the axion-inflaton and, in turn, to the curvature power spectrum. The existence of such instability is most clearly seen following the WKB approximation analysis of \cite{Adshead:2013nka}. In there it is shown that, in the $\Lambda\gg1$ regime, as soon as $m_Q$ enters the instability band one of the frequency of the scalar modes becomes imaginary  already within the horizon.\\
\indent Studies so far have have explored regions of parameter space supporting an inflationary evolution that never enters the instability band.~However, the very recent findings of \cite{Iarygina:2023mtj} compel us to investigate this very regime. One of the main results of  \cite{Iarygina:2023mtj} is that strong backreaction of gauge field fluctuations on the $Q,\chi$ backgrounds forces $m_Q$ to enter the instability band, cross zero, and  exit the instability towards the negative attractor solution $\lambda Q \dot{\chi}/f=-1$ (see Fig.~\ref{mQold}). We note that large $m_Q$ values are  (i) those that will trigger strong backreaction \footnote{Strong backreaction is typically active for $m_Q\gtrsim 4$.} and thus (ii) those leading to a finite crossing of the instability band.\\

\indent In the case of the axion-inflaton, it can be shown that  $m_Q$ generically enters the instability band. One need only remember that outside the strong backreaction regime the $m_Q$ time evolution is well approximated by \footnote{Eq.~(\ref{Qdep}) is valid for chromo natural inflation (CNI). On the other hand, one can straightforwardly verify that the same $\sqrt{\epsilon_{H}}/H$ time dependence of Eq.~(\ref{Qdep}), with an overall exponent $1$ rather than $1/2$, holds also e.g. in the case of non-minimally coupled chromo \cite{Dimastrogiovanni:2023oid}. An $m_Q$ growing into the strong backreaction regime is then a generic feature of an axion-inflaton coupled via CS term to non-Abelian fields.}
\bea
m_Q\simeq\left(g\sqrt{\epsilon_H}\frac{M_p}{H}\right)^{1/2}\;.
\label{Qdep}
\eea
An increasing $\epsilon_{H}$ and decreasing $H$ over time will generically push $m_Q$ from the threshold $\sqrt{2}$ value to $m_Q\simeq 4$ that typically  triggers strong backreaction and leads to the crossing into the instability  band studied in \cite{Iarygina:2023mtj}.

The work \cite{Iarygina:2023mtj} is focused on the spectator-CNI model \cite{Dimastrogiovanni:2016fuu}. We show here that the intriguing mechanism uncovered in \cite{Iarygina:2023mtj} applies in a wider context and is a phenomenon that is more generic to the axion-inflaton (i.e. non-spectator) case. We have explicitly verified,  lifting also their de Sitter approximation, that the results of \cite{Iarygina:2023mtj} are qualitatively independent from the specific  axion-inflaton potential. The key features of the mechanism rely solely on there being a rolling inflaton that dissipates energy into the gauge sector via CS coupling. We will now show how the $m_Q$  evolution into the instability band leads to PBH production.

\section{Crossing the Instability and Primordial Black Holes production}
\label{crossing}

\begin{figure}
    \centering
    \includegraphics[width=0.99\linewidth,angle=0]{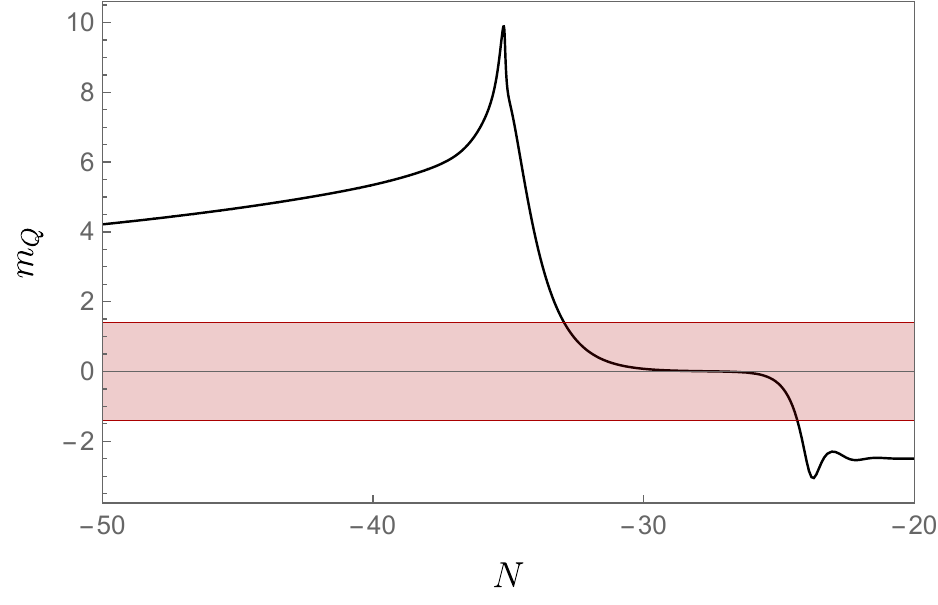}
    \caption{A typical example of the evolution of the particle production parameter $m_Q$ in chromo-natural inflation with entry to the strong backreaction regime and inevitable crossing into the $|m_Q|<\sqrt{2}$ instability band. The parameters for this run are $g=0.002,\;\lambda=300,\;f=0.4\,M_p,\;\mu=1.47\cdot 10^{-3}\, M_p$.}
    \label{mQaxinf}
\end{figure}

\begin{figure}
    \centering
    \includegraphics[width=0.99\linewidth,angle=0]{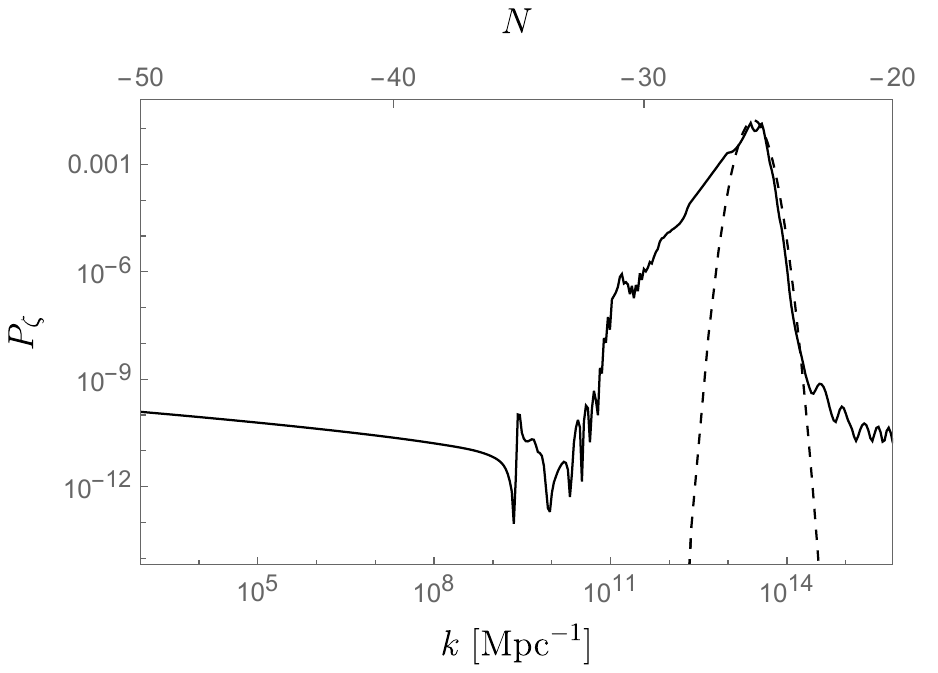}
    \caption{The power spectrum of scalar perturbations corresponding to the example in Fig.~\ref{mQaxinf}. We consider only linear effects and disregard the sourcing of $\zeta$ from the scalar perturbations of the gauge field which, as we have verified, yield a subdominant contribution. Further comments on the shape of the peak can be found in the supplemental material.}
    \label{Pzetaaxinf}
\end{figure}

Primordial black holes can be generated in the early universe as a result of a large curvature $\zeta$ power spectrum. Perturbations will re-enter the horizon after inflation and form PBHs \cite{Ivanov:1994pa,Garcia-Bellido:1996mdl,Ivanov:1997ia}. The crossing of the instability band by scalar perturbations outlined in Section \ref{section2}  feeds into 
the axion-inflaton fluctuations and ultimately into the  $\zeta$ power spectrum \footnote{Scalar fluctuations in the gauge sector are much smaller than those of the axion so that their contribution to $\zeta$ can be neglected.}. This PBH production mechanism is largely independent of the axion-inflaton potential. We shall employ here the standard cosine potential $V=\mu^4[1+{\rm Cos}(\chi/f)]$ but we stress that the dynamics would be qualitatively identical for a generic choice.~On the other hand, ensuring agreement with CMB observations will be more taxing on the chosen potential; we leave this to future work \cite{Dimastrogiovanni:future}. \\
\indent We plot in Fig.~\ref{mQaxinf} the quantity $m_Q$ as a function of e-folds and in Fig.~\ref{Pzetaaxinf} the corresponding curvature power spectrum.  The values chosen for the key parameters $\lambda,g$ as well as those for $f,\mu$, are in the standard range for axion-inflation models coupled to gauge fields and are listed in the caption of Fig.~\ref{mQaxinf}. The striking enhancement highlighted by Fig.~\ref{Pzetaaxinf} is due to two main effects. This is most easily seen by looking at the expression for the curvature perturbation:
\begin{eqnarray}
\zeta=\frac{V_{\chi}(\chi)}{6\epsilon_H H^{2}M_{P}^{2}}\,\delta\chi\; .
\label{pzetadef}
\end{eqnarray}

First, as outlined above, crossing the instability band enhances the axion-inflaton perturbation $\delta \chi$. We verified this mechanism is particularly efficient shortly after $m_Q$ enters and shortly before it exits the instability band whilst it is far less effective in the plateau region (see Fig.~\ref{mQaxinf}). The latter interval corresponds to small values of $\Lambda$, whilst it is the $\Lambda\gg1$ regime that most fuels the $\delta\chi$ enhancement. 
The second handle on enhancing the curvature power spectrum relies on a decreasing $\epsilon_{H}$. Such effect is brought about by strong backreaction and does not require access to the instability band. One can see in Fig.~\ref{epsplot} in Appendix~\ref{app.B} that indeed $\epsilon_{H}$ starts noticeably varying already at the onset of strong backreaction (hence much before the time when $m_Q\leq \sqrt{2}$) and its value soon becomes suppressed due entirely to backreaction contributions.  

The scalar power spectrum in Fig.~\ref{Pzetaaxinf} is quite intriguing in that it leads to sufficient PBH production at scales where these may constitute a very significant fraction (if not the entirety) of the dark matter in the universe. We must stress here that, although one may easily arrive at key observables ($P^{\rm cmb}_{\zeta}, n_s$) within the same order of magnitude of the measured values, a thorough analysis at CMB scales is necessary to strengthen bolder claims in this context. We find reasons to be quite optimistic in this respect in light of the insensitivity of the PBH production mechanism to the axion potential. The latter can then be chosen mostly on the basis of its ability to satisfy CMB constraints. We leave this to future work \cite{Dimastrogiovanni:future}.

\begin{figure}[h!]
    \centering
    \includegraphics[width=0.99\linewidth,angle=0]{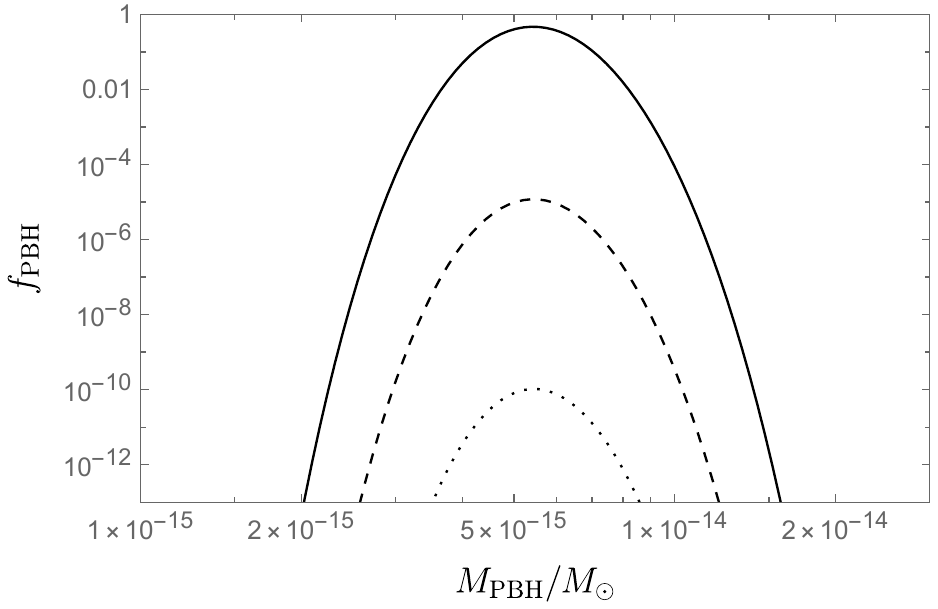}
    \caption{The value of the black hole mass fraction corresponding to the power spectrum in Fig.~\ref{Pzetaaxinf}. As reference we display the mass fraction for critical density values $\delta_c=0.34,\; 0.31, \;0.28$ in dotted, dashed and solid lines respectively.}
    \label{fPBH}
\end{figure}

In Fig.~\ref{fPBH} we plot the PBH abundance as a function of the black hole mass $M_k$. Our calculation follows closely that of \cite{Dalianis:2018frf} and is based on a number of assumptions.
First, we consider PBH formation in the radiation era. We use the Press-Schechter formalism and assume that curvature perturbations are Gaussian. A significant non-Gaussianity would lead to corrections that depend on the sign and the specific shape \footnote{Previous studies on axion models \cite{Agrawal:2017awz,Dimastrogiovanni:2018xnn,Fujita:2018vmv} point to a nearly equilateral shape.} of the scalar bispectrum, see e.g. \cite{Byrnes:2012yx,Lyth:2012yp,Young:2013oia,Franciolini:2018vbk,Matsubara:2022nbr,Firouzjahi:2023xke}. We do not include these corrections here also in light of the exponential sensitivity of the final PBH abundance result to the threshold density perturbation $\delta_c$.~The effect of different values for $\delta_c$ far overshadows non-Gaussianity corrections. 

\begin{figure}[h!]
    \centering
    \includegraphics[width=0.99\linewidth,angle=0]{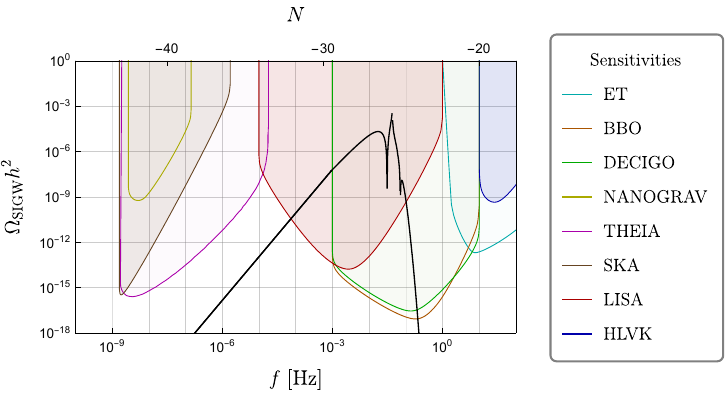}
    \caption{We present the SIGW signal corresponding to the power spectrum of Fig.~\ref{Pzetaaxinf}. For simplicity we made use of the analytical results of \cite{Pi:2020otn} for the narrow peak approximation and fit our $P_\zeta$ with the log-normal approximation shown in the black dashed line in Fig.~\ref{Pzetaaxinf}.}
    \label{SIGW}
\end{figure}

Now, crucially, the very same $P_{\zeta}$ responsible for PBH production will non-linearly source primordial gravitational waves, giving rise to the well-known scalar-induced contribution \cite{Matarrese:1997ay,Ananda:2006af,Baumann:2007zm,Saito:2009jt,Domenech:2021ztg}. There are only very few windows left for $f_{\rm PBH}$ that are observationally viable and may correspond to a very significant dark matter fraction. Fascinatingly, the corresponding SIGW happens to be in the LISA frequency range \cite{Bartolo:2018evs}. We plot in Fig.~\ref{SIGW} the second-order gravitational wave due to scalar fluctuations superimposed with the sensitivity bands of several GW detectors. 

\indent It is important to stress at this stage that tensor fluctuations in the gauge sector, the very same responsible for strong backreaction, will also source gravitational waves linearly, further contributing to the GW signal and possibly enriching its peak structure. However, we neglect this term for the moment as it is expected \cite{Iarygina:2023mtj} to be orders of magnitude smaller than the SIGW found here.

\section{The spectator case}
\label{spectator}

\begin{figure}[h!]
    \includegraphics[width=6.5cm]{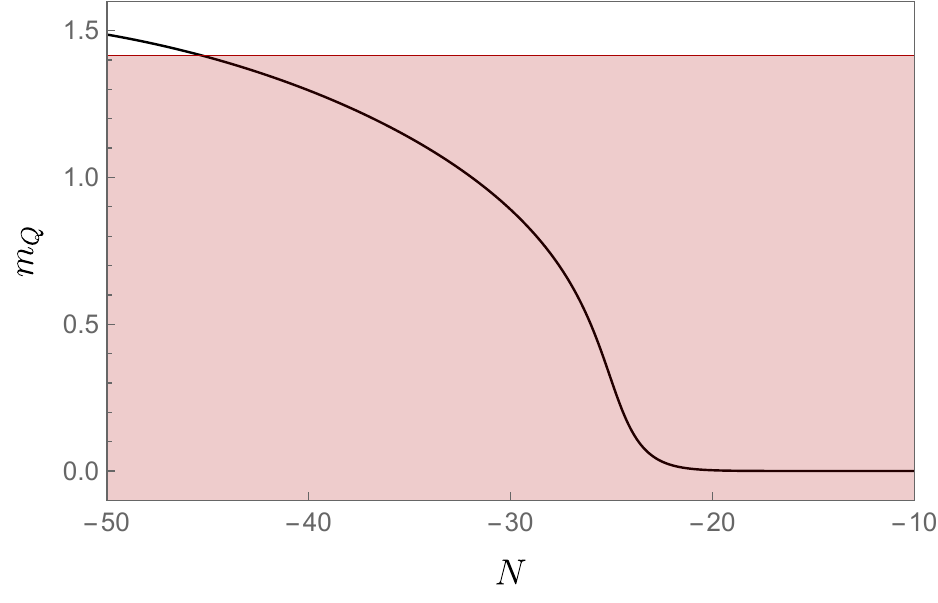}
    \caption{A decaying axion corresponds to an   $m_Q$ decreasing with time, entering the instability band before the end of inflation. Choosing an initial $m_Q$ such that strong (gauge field-) tensor backreaction is never activated significantly restricts the parameter space scanned.}
    \label{fig:mQspec}
\end{figure}

We consider a simple extension of the model in Eq.~(\ref{Lag}) whereby a scalar field $\phi$ is added and asked to drive the acceleration, thus becoming the inflaton  \cite{Dimastrogiovanni:2016fuu}. The ALP is now just a rolling spectator field. Multi-field scenarios are well motivated during inflation \cite{Baumann:2014nda}. We shall focus on the case when strong backreaction is not triggered (see Fig.~\ref{fig:mQspec}): if this were instead the case, the dynamics would be entirely analogous to the case analyzed in the previous section in light of the existence of an attractor solution for $Q$. 

\begin{figure}[h!]
    \centering
    \includegraphics[width=0.7\linewidth,angle=0]{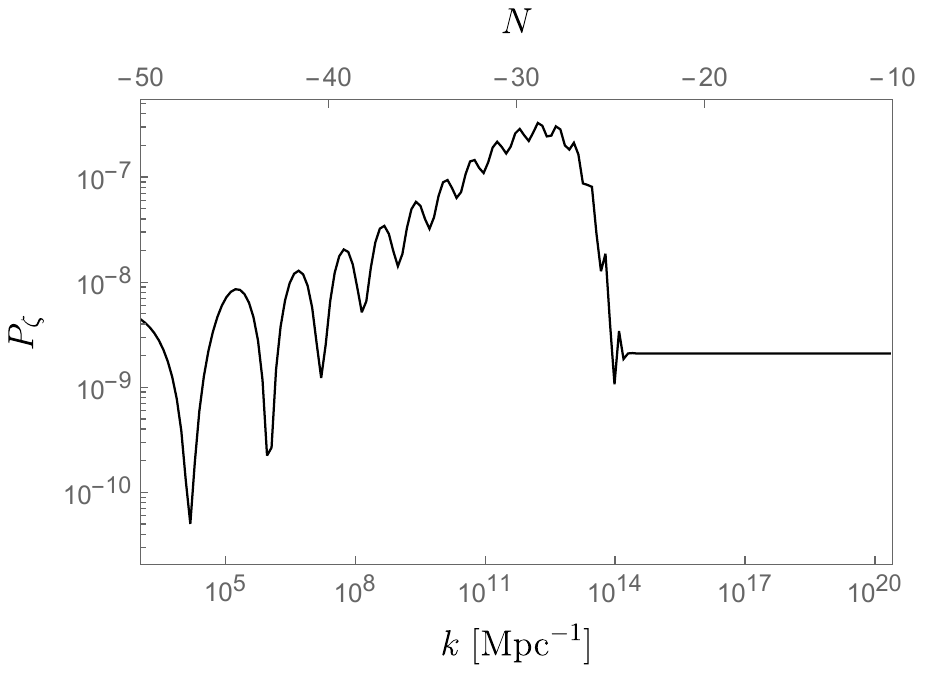}
    \caption{The scalar curvature power spectrum in the spectator case. The oscillations are likely due to the WKB approximation scheme used.~Given the small values of the spectrum amplitude, we do not pursue more sophisticated techniques in this case.    
}
    \label{Pzetaspec}
\end{figure}

\indent Another working assumption is that the ALP decays before the end of inflation. This means the axion fluctuations effectively contribute to the curvature power spectrum only indirectly, i.e. through their contribution to the inflaton fluctuations. Such choice simplifies \footnote{This is a rather consequential assumption. We will show that, in order to produce a significant PBH fraction in this context, a very large hierarchy is necessary between axion and curvature fluctuations. This will in turn imply large scalar backreaction and make PBH production markedly less efficient. The case of direct axion contribution to $\zeta$, and therefore of an ALP decaying with or after the inflaton, leaves the door open to a more fertile PBH production rate. We leave exploring this possibility to forthcoming work.} both our analysis here as well as the  re-heating process.\\ \indent In calculating the $\chi$ indirect contribution to curvature perturbations one goes through two simple steps $\zeta=-H\delta\phi/\dot{\phi}$, and  $\delta\phi\sim\sqrt{\epsilon_{\phi}\epsilon_{B}}\, \delta\chi $, both of which account for a suppression factor of several orders of magnitude. It follows that the  $P_{\zeta}$  necessary to account for a significant $f_{\rm PBH}$ fraction  requires a tremendous hierarchy between $\zeta$ and $\delta\chi$. This comes at a price: the strong backreaction of scalar axion fluctuations on (at the very least \footnote{The other background equation affected being that of $Q$. See supplemental material for the explicit expressions.}) its background equations of motion. In this work we push the $\zeta-\delta\chi$ hierarchy up to the onset of strong scalar backreaction, but no further. We find that the corresponding curvature power spectrum is not sufficiently large to grant a significant PBH production, see Fig.~\ref{Pzetaspec}.

Our analysis of the axion spectator case is admittedly not exhaustive. Nevertheless, it certainly suffices to make clear the point that the PBH production mechanism is less universal than in the axion-inflaton case. The reason is intuitively clear: the rolling of a spectator ALP is subject to fewer constraints than that of the inflaton. It may, for example, not last for the whole duration of inflation so that the ALP can decay and  source curvature fluctuations only indirectly, i.e. (typically) less efficiently. 

\section{Conclusions}
\label{conclusions}

In this work we explored for the first time scalar fluctuations in the instability band that arises when axion inflation models are coupled with a non-Abelian gauge sector.
Induced through the CS term, the instability turns out to be a controllable mechanism exhibiting  a rich and interesting phenomenology. We find that excursions through the instability band give rise to striking enhancements of scalar fluctuations without spoiling compliance with observational bounds.
As a result, our work opens up a new intriguing venue in the study of primordial black hole production. The case of the axion-inflaton stands out in this context.~Indeed, even when initiating the system in the weak backreaction regime,  strong backreaction is arrived at dynamically so that the crossing of the instability band is triggered. This new PBH production mechanism is largely independent from the specific shape of the potential, making the phenomenon universal.

We also investigated the axion-spectator case. The ALP dynamics is no longer  subject to the constraints of the inflaton, thus allowing for instability band crossings whilst still in the weak-backreaction configuration. We highlighted a much reduced scalar curvature enhancement when the ALP decays before the end of inflation under the additional assumption of small scalar backreaction. This suggests extending our work in manifold directions. Firstly, one would explore lifting  the weak scalar backreaction requirement. Secondly, it will be intriguing to scrutinize the case of ALP that decays at or after the end of inflation, including the possibility of isocurvature modes. Another interesting possibility is to consider features in the potential that may lead to controlled excursion in the instability band. We leave  these endeavors to future work.

\section{Acknowledgments}
\noindent  MF and AP  acknowledge the ``Consolidación Investigadora'' grant CNS2022-135590. Their work is partially supported by the Spanish Research Agency (Agencia Estatal de Investigación) through the Grant IFT Centro de Excelencia Severo Ochoa No CEX2020-001007-S, funded by MCIN/AEI/10.13039/501100011033. MF acknowledges support also from the
``Ramón y Cajal'' grant RYC2021-033786-I.

\appendix

\section{Equations used for the numerical evolution}
\label{app.A}

In this Appendix we report the explicit  equations being used to derive the results displayed in the main text. We present the equations for the spectator chromo-natural inflation case. From there one can easily obtain the equations for the standard CNI model by setting the field $\phi$ to zero everywhere and having instead the field $\chi$ drive the acceleration.

According to our convention  the symbol $V$ is always reserved for the inflaton potential. We begin with the background equations of motion in physical time
\begin{align}
3 H^{2} M_{p}^{2}=&\frac{\dot{\phi}^{2}}{2}+V(\phi)+\frac{\dot{\chi}^{2}}{2}+U(\chi)+\frac{3}{2}\Big[\left(\dot{Q}+HQ \right)^{2}\nonumber\\&\quad\quad\quad\quad \quad\quad\quad\quad \quad\quad +g^{2}Q^{4}\Big]+\rho_t\,,\\
-2M_{p}^{2}\dot{H}=&\dot{\phi}^{2}+\dot{\chi}^{2}+2\left[\left(\dot{Q}+HQ \right)^{2}+g^{2}Q^{4}\right]\nonumber\\& \quad\quad\quad\quad \quad\quad\quad\quad \quad +Q \,{\cal T}^Q_{BR}+\frac{4}{3}\rho_t\,,\label{eq:Hdot}\\
\ddot{\chi}+3H\dot{\chi}&+U_{\chi}+\frac{3 g \lambda}{f}Q^{2}\left(\dot{Q}+H Q\right)+\mathcal{T}^{\chi}_{BR}=0\,,\\
\ddot{Q}+3H\dot{Q}&+\left(\dot{H}+2H^{2} \right)Q+gQ^{2}\left(2gQ-\frac{\lambda \dot{\chi}}{f} \right)\nonumber\\& \quad\quad\quad\quad \quad\quad\quad\quad \quad+\mathcal{T}^{Q}_{BR}=0\,,\label{eq:eqQ}
\end{align}
with
\begin{align}
&&\mathcal{T}^{\chi}_{BR}=-\frac{\lambda}{2 a^{2}f}\frac{d}{dt}\int\frac{d^{3}k}{(2\pi)^{3}}\left(m_{Q}H-\frac{k}{a} \right)| t_{L} |^{2}\,, \label{eq:Tchi}\\
&&\mathcal{T}^{Q}_{BR}= \frac{g}{3 a^{2}}\int\frac{d^{3}k}{(2\pi)^{3}}\left(\xi H-\frac{k}{a} \right) | t_{L} |^{2} \,, \label{eq:TQ}\\
&&\rho_t=\frac{1}{2a^2}\int\frac{d^{3}k}{(2\pi)^{3}}\left[| \dot{t}_{L} |^{2}+\left(\frac{k^2}{a^2}-2g Q \frac{k}{a}\right)| t_{L} |^{2}\right]\;.\label{eq:rhot}
\end{align}
One interesting thing to note is that the usual expression for the backreaction of the tensor modes on the equation of motion of the gauge field vev, ${\cal T}^Q_{BR}$, also directly contributes to the equation for $\dot{H}$ (\ref{eq:Hdot}). This is a consequence of the presence of $\ddot{Q}$ in the second Einstein equation which, when eliminated by making use of (\ref{eq:eqQ}), results in the emergence of the backreaction term directly in the background geometry.

Concurrently with evolving the background, we are also evolving the tensor perturbations which then affect the background evolution through (\ref{eq:Tchi}), (\ref{eq:TQ}) and (\ref{eq:rhot}). In practice, the last backreaction formula can be eliminated using the first Friedmann equation, and in regards to the background we are only solving the second Friedmann equation, with the first one providing the initial condition and eliminating $\rho_t$. 

Following the decomposition of the perturbations as in  \cite{Papageorgiou:2019ecb}, aligning the momentum along the z-axis, the tensor modes of the gauge fields are given by
\begin{align} 
	 \delta A_\mu^1 = a  \left( 0 ,\, t_+  ,\, t_\times  ,\, 0 \right)\,,\;\; 
	\delta A_\mu^2 = a  \left( 0 ,\, t_\times  ,\, - t_+  ,\, 0 \right) \,.\nonumber\\ 
\end{align}
Upon introducing the canonically-normalised left and right-handed helicity modes
\begin{equation}
t_{R,L}  \equiv a\left(t_+  \pm i t_\times\right) \,,
\end{equation} 
the homogeneous equation for these modes reads \cite{Dimastrogiovanni:2016fuu}:
\begin{eqnarray}
t_{R,L}^{''}+\left[1+\frac{2}{x^{2}}\left(m_{Q}\xi\pm x\left(m_{Q}+\xi \right) \right) \right]t_{R,L}=0\,.
\end{eqnarray}
For the purposes of this work, we disregard the inhomogeneous terms which depend on the tensor metric fluctuations. These contributions are generally subdominant in the regime we are interested in. The tensor equations of motion written above are exact, without any implicit slow-roll approximation.

Let us now turn our attention to the scalar sector. The scalar perturbations of the gauge field can be written as \cite{Papageorgiou:2019ecb}:
\begin{eqnarray}
	& & \delta A_\mu^1 = \left( 0 ,\; \delta \varphi  - Z  ,\; \chi_3  ,\; 0 \right) \;, \nonumber\\ 
	& & \delta A_\mu^2 = \left( 0 ,\; - \chi_3  ,\; \delta \varphi  - Z   ,\; 0 \right) \;, \nonumber\\ 
	& & \delta A_\mu^3 = \left( \delta A_0^3  ,\; 0 ,\; 0 ,\; \delta \varphi + 2 Z  \right)\,, 
\end{eqnarray} 
where $\delta A_0^3$ is a non-dynamical mode that will be integrated out of the equations, and
\begin{equation}
\chi_3 = - \partial_z \frac{2 Z+ \delta \varphi}{2 g a Q}\,.
\end{equation}
We report below the evolution equations for the scalar modes of the full system. In the axion-spectator case these are four: two from the axion and inflaton fields,
\begin{eqnarray}
\hat{\Phi}=a\,\delta\phi\,,\quad\quad\hat{X}\equiv a \,\delta\chi\,,
\end{eqnarray}
and two, $\hat{Z}$ and $\hat{\varphi}$, from the gauge field:
\begin{align}
	 {\hat Z} \equiv \sqrt{2} \left( Z - \delta \varphi \right)\,, \;\;\;\;
	{\hat \varphi} \equiv \sqrt{2 + \frac{x^2}{m_Q^2}} \left( \frac{\delta \varphi}{\sqrt{2}} + \sqrt{2} \, Z \right) \, .
\end{align} 
The evolution equations read:
\begin{align}
\hat{\Phi}^{''}+\left(1-\frac{2}{x^2} \right)\hat{\Phi}+\frac{\sqrt{2}H m_{Q}^{4}\sqrt{\epsilon_{\phi}}\,\Lambda}{g M_p\left(2\,m_{Q}^{2}+x^{2}\right)}\frac{\hat{X}^{'}}{x}\nonumber\\-\frac{\sqrt{2}H m_{Q}^{2}\sqrt{\epsilon_{\phi}}\,\Lambda\left(10 m_{Q}^{4}+9 m_{Q}^{2}x^{2}+x^{4}\right)}{g M_{p}\left(2 m_{Q}^{2}+x^{2} \right)^{2}}\frac{\hat{X}}{x^2}=0\, ,
\label{eq:phichi}
\end{align}
\begin{align}
\hat{X}''+\left(1-\frac{2}{x^2}+\frac{\Lambda ^2 m_Q^2}{2 m_Q^2+x^2}\right)\hat{X} -\frac{\Lambda\, m_Q^2}{x^2 \sqrt{\frac{1}{2}+\frac{m_Q^2}{x^2}}}\hat{\varphi}'\nonumber\\+\frac{\sqrt{2} \Lambda \, \left(4 m_Q^4+3 m_Q^2 x^2+x^4\right)}{x^2 \left(2 m_Q^2+x^2\right)^{3/2}}\hat\varphi+\frac{\sqrt{2}\, \Lambda\,  m_Q}{x}\hat{Z}'\nonumber\\-\frac{2 \sqrt{2} \,\Lambda \, m_Q}{x^2}\hat{Z}=0\, ,
\end{align}
\begin{align}
\hat{\varphi}''+\frac{\left(1+\frac{8 m_Q^6}{x^6}+\frac{2 m_Q^2 \left(8 m_Q^2-2 m_Q \xi +3\right)}{x^4}+\frac{2 m_Q (4 m_Q-\xi )}{x^2}\right)}{\left(1+\frac{2 m_Q^2}{x^2}\right)^2}\hat{\varphi}\nonumber\\+\frac{\Lambda\,  m_Q^2}{x^2 \sqrt{\frac{m_Q^2}{x^2}+\frac{1}{2}}}\hat{X}'+\frac{\sqrt{2} \Lambda \, \left(2 m_Q^4+m_Q^2 x^2+x^4\right)}{x^2 \left(2 m_Q^2+x^2\right)^{3/2}}\hat{X}\nonumber\\-\frac{2 (m_Q-\xi) \sqrt{2 m_Q^2+x^2}}{x^2}\hat{Z}=0
\end{align}
\begin{align}
\hat{Z}''+\frac{\left(4 m_Q^2-2 m_Q \xi +x^2\right)}{x^2}\hat{Z}-\frac{\sqrt{2} \,\Lambda\,  m_Q}{x}\hat{X}'\nonumber\\-\frac{\sqrt{2}\, \Lambda  \,m_Q}{x^2}\hat{X}-\frac{2 (m_Q-\xi ) \sqrt{2 m_Q^2+x^2}}{x^2}\hat{\varphi}=0\, ,
\end{align}
where we have  integrated out the non-dynamical modes of the metric and gauge field simultaneously according to the standard procedure. Here the prime indicates a derivative w.r.t.~the variable $x\equiv -k\eta$ ($\eta$ being conformal time) and the terms that are higher-order in slow-roll have been suppressed. 
The exact derivation of the $\hat{X}$, $\hat{\Phi}$ coupling can be found in \cite{Papageorgiou:2019ecb}. 

\section{Comments on the numerical scheme and results}
\label{app.B}

We perform our numerical evolution in two stages. First, we solve the background equations simultaneously with the tensor equations of motion for a set of modes that is uniformly distributed in logarithmic $k$-space. We choose ${\cal O}(100)$ set of modes, appropriately chosen so that they have appropriate coverage of the growth of perturbations around horizon crossing when the backreaction becomes large. At every time step we reconstruct the backreaction by discretising the backreaction integrals and summing the various contributions using the trapezoid rule. In general terms, the numerical strategy is similar to the one employed in \cite{Garcia-Bellido:2023ser} with appropriate changes/generalizations to fit the more complex equations of the axion-non-Abelian model. Our code is written in \textit{Julia}.

As briefly mentioned above, we treat the chromo-natural inflation model in a different way to the spectator one. In the first case, we fully evolve the background, set the field $\phi$ to zero and assume the field $\chi$ to be the inflaton, while in the latter case we do not evolve the inflaton or the background, assuming instead a perfect de Sitter background. Our time variable for all the code evolutions is the number of e-foldings defined as $N\equiv \log{a}$, with $a_{\rm in}\equiv 1$.

\begin{figure*}
    \centering
    \includegraphics[width=0.54\linewidth,angle=0]{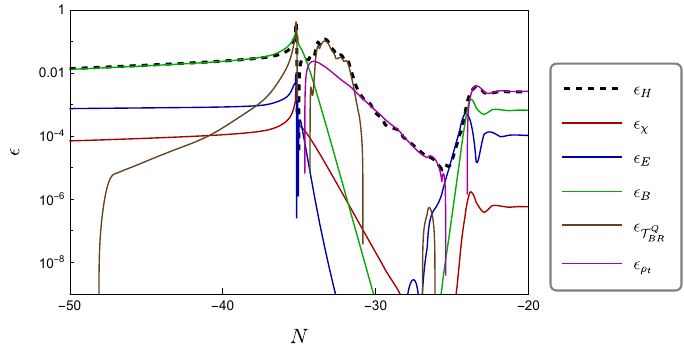}
    \includegraphics[width=0.43\linewidth,angle=0]{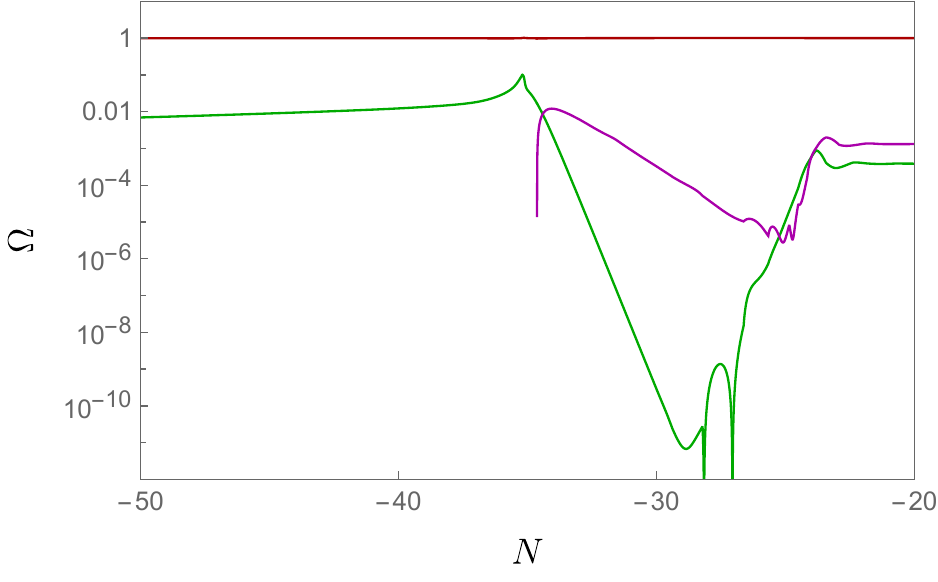}
    \caption{The left panel displays the various contributions to the Hubble slow roll parameter $\epsilon_H$ shown in the black dashed line for the same example shown in Figs.~\ref{mQaxinf} and~\ref{Pzetaaxinf}. One may observe that at early times $\epsilon_B$ dominates the slow-roll parameter as is usual in chromo-natural inflation, however, at the onset of the strong backreaction, the dominant contribution varies in the following sequence $\epsilon_{{\cal T}^Q_{BR}}\rightarrow \epsilon_{\rho_t} \rightarrow \epsilon_E \rightarrow \epsilon_{\rho_t}$. The right panel is useful in verifying that the energy of the inflaton is indeed the dominant component both before and after entering the strong backreaction regime as well as during the transition period.}
    \label{epsplot}
\end{figure*}
We have verified that our code produces the same results with different integrators, and results which converge for sufficiently low error tolerances. A typical example of the evolution of the background for the axion-inflaton case is shown in Fig.~\ref{epsplot}. The slow-roll parameters are defined as:
\begin{eqnarray}
    \epsilon_H\equiv -\frac{\dot{H}}{H^2}=\epsilon_\chi+\epsilon_B+\epsilon_E+\epsilon_{{\cal T}_Q}+\epsilon_{\rho_t}
\end{eqnarray}
with
\begin{align}
    \epsilon_\chi\equiv \frac{\dot{\chi}^2}{2M_p^2 H^2}\,,&\;\;\epsilon_B\equiv \frac{g^2Q^4}{M_p^2 H^2}\,,\;\;\epsilon_E\equiv \frac{\left(\dot{Q}+H Q\right)^2}{M_p^2 H^2}\,,\nonumber\\
    \epsilon_{{\cal T}^Q_{BR}}\equiv&\frac{Q}{2 M_p^2 H^2}{\cal T}^Q_{BR}\,,\;\;\epsilon_{\rho_t}\equiv\frac{2}{3 M_p^2 H^2}\rho_t\, .
\end{align}
One can observe that when the system enters the strong-backreaction regime at around $35$ e-folds before the end of inflation, the slow-roll parameter $\epsilon_H$ decreases by a few orders of magnitude. This decrease in turn manifests as a peak in the power spectrum $\P_\zeta$ of Fig.~\ref{Pzetaaxinf} due to the presence of the slow-roll parameter in the denominator of (\ref{pzetadef}). This feature of the slow-roll parameter gives the power spectrum its characteristic tilt. On the other hand, the canonical perturbation $\hat{X}$ increases by about an order of magnitude with respect to its vacuum configuration only when the particle production parameter $m_Q$ is just entering the instability regime and exiting it, since that is when $\Lambda$ is still at least order one. In the intermediate plateau, the instability becomes very inefficient due to the smallness of the $\Lambda$ parameter. This overall leaves open the possibility for other parameter choices which would yield a natural and well correlated two-peak structure in the power spectrum. We leave the full exploration of this interesting part of the parameter space to future work. Finally, the overall strength of the signal is very sensitive to the value of parameter $f$, with smaller values yielding a greater signal while leaving the background evolution invariant. 

In the spectator case, we identify a regime where the strong enhancement of axion perturbations  may backreact  on the background equations of motion. The scalar backreaction terms in question for the $\chi,Q$ equations of motion read

\begin{eqnarray}
    B^\chi_{BR}&=&\frac{V^{(3)}(\chi)}{2 a^2}\int\frac{d^3k}{(2\pi)^3}\left|\hat{X}(\tau,k)\right|^2\, ,\label{eq:scalarbackreaction1}\\
    B^Q_{BR}&=&\frac{2 g \Lambda^2 m_Q}{3a^2}\int\frac{d^3k}{(2\pi)^3}\frac{k^2\left(k^2+a^2 H^2 m_Q^2\right)}{\left(k^2+2 a^2 H^2 m_Q^2\right)^2}\left|\hat{X}(\tau,k)\right|^2\, .\label{eq:scalarbackreaction2}\nonumber\\
\end{eqnarray}

For the purposes of this work we limit our analysis to the configuration where the first of the two expressions is no greater than the slope of the axion potential. This  places significant restrictions on the sourcing of the inflaton perturbations from the coupling in Eq.~(\ref{eq:phichi}). While Eq.~(\ref{eq:scalarbackreaction2}) is characteristic of the chromo-natural type models of inflation, Eq.~(\ref{eq:scalarbackreaction1}) is a general expression that applies to all spectator scalar field perturbations. We found clear evidence that all models in which spectator scalars source the inflaton perturbation through a slow-roll suppressed coupling as in (\ref{eq:phichi}) are vulnerable to violating the low backreaction conditions due to the hierarchy required between the spectator and inflaton perturbations in order to produce an observable signal. It will be interesting to return to this point in future work.

\bibliographystyle{apsrev4-1}
\bibliography{ref}

\end{document}